\documentclass[sigconf]{acmart}

\usepackage{xspace} 

\newcommand{\nIDEXandEDStartDate}{02/09/2017\xspace}
\newcommand{\nIDEXandEDEndDate}{05/04/2020\xspace}

\newcommand{\nIDEXStartDate}{09/27/2017 10:57pm\xspace}
\newcommand{\nIDEXEndDate}{05/04/2020 1:22pm\xspace}
\newcommand{\nEtherDeltaStartDate}{02/09/2017 11:56pm\xspace}
\newcommand{\nEtherDeltaEndDate}{05/04/2020 1:22pm\xspace}

\newcommand{\nIDEXTradesRemoved}{\ensuremath{3,612}\xspace}
\newcommand{\nEtherDeltaTradesRemoved}{\ensuremath{3,095}\xspace}

\newcommand{\nIDEXTradesPreprocessed}{\ensuremath{5,340,537}\xspace} %
\newcommand{\nIDEXAccountsPreprocessed}{\ensuremath{249,911}\xspace} %
\newcommand{\nIDEXTokensPreprocessed}{\ensuremath{1,206}\xspace} %
\newcommand{\nIDEXSelfTrades}{\ensuremath{28,193}\xspace} %
\newcommand{\nIDEXSelfTradeAccounts}{\ensuremath{332}\xspace} %
\newcommand{\nIDEXSelfTokens}{\ensuremath{239}\xspace} %
\newcommand{\shareIDEXSelfTrades}{\ensuremath{0.53\%}\xspace} %
\newcommand{\totalIDEXSelfTradesETH}{\ensuremath{23,716.08}\xspace} %
\newcommand{\totalIDEXSelfTradesUSD}{\ensuremath{12,587,517}\xspace} %
\newcommand{\nIDEXWashTrades}{\ensuremath{213,029}\xspace} %
\newcommand{\shareIDEXWashTrades}{\ensuremath{3.99\%}\xspace} %
\newcommand{\totalIDEXWashVolumeETH}{\ensuremath{273,027}\xspace} %
\newcommand{\totalIDEXWashVolumeUSD}{\ensuremath{83,531,254}\xspace} %
\newcommand{\nIDEXWashTokens}{\ensuremath{380}\xspace} %
\newcommand{\shareIDEXWashTokens}{\ensuremath{31.54\%}\xspace} %
\newcommand{\nIDEXWashTradeAccounts}{\ensuremath{659}\xspace} %
\newcommand{\nIDEXAnalyzedSCC}{\ensuremath{199}\xspace} %
\newcommand{\nIDEXWashSCC}{\ensuremath{193}\xspace} %
\newcommand{\nIDEXMeanWashedTokenPerSCC}{\ensuremath{2.46}\xspace} %

\newcommand{\nEtherDeltaTradesPreprocessed}{\ensuremath{3,573,512}\xspace} %
\newcommand{\nEtherDeltaAccountsPreprocessed}{\ensuremath{323,598}\xspace} %
\newcommand{\nEtherDeltaTokensPreprocessed}{\ensuremath{6,551}\xspace} %
\newcommand{\nEtherDeltaSelfTrades}{\ensuremath{58,791}\xspace} %
\newcommand{\nEtherDeltaSelfTradeAccounts}{\ensuremath{5,501}\xspace} %
\newcommand{\nEtherDeltaSelfTokens}{\ensuremath{2,758}\xspace} %
\newcommand{\shareEtherDeltaSelfTrades}{\ensuremath{1.09\%}\xspace} %
\newcommand{\totalEtherDeltaSelfTradesETH}{\ensuremath{136,548.22}\xspace} %
\newcommand{\totalEtherDeltaSelfTradesUSD}{\ensuremath{64,801,288}\xspace} %
\newcommand{\nEtherDeltaWashTrades}{\ensuremath{69,711}\xspace} %
\newcommand{\shareEtherDeltaWashTrades}{\ensuremath{1.30\%}\xspace} %
\newcommand{\totalEtherDeltaWashVolumeETH}{\ensuremath{155,155}\xspace} %
\newcommand{\totalEtherDeltaWashVolumeUSD}{\ensuremath{75,846,518}\xspace} %
\newcommand{\nEtherDeltaWashTokens}{\ensuremath{2,759}\xspace} %
\newcommand{\shareEtherDeltaWashTokens}{\ensuremath{42.12\%}\xspace} %
\newcommand{\nEtherDeltaWashTradeAccounts}{\ensuremath{5,533}\xspace} %
\newcommand{\nEtherDeltaAnalyzedSCC}{\ensuremath{48}\xspace} %
\newcommand{\nEtherDeltaWashSCC}{\ensuremath{47}\xspace} %
\newcommand{\nEtherDeltaMeanWashedTokenPerSCC}{\ensuremath{1.62}\xspace} %

\newcommand{\ntotalWashTradingVolumeShort}{\$159 million\xspace}

\usepackage{caption}
\usepackage{subcaption}
\usepackage[ruled,vlined,linesnumbered]{algorithm2e}

\AtBeginDocument{%
  \providecommand\BibTeX{{%
    \normalfont B\kern-0.5em{\scshape i\kern-0.25em b}\kern-0.8em\TeX}}}

\copyrightyear{2021}
\acmYear{2021}
\setcopyright{iw3c2w3}
\acmConference[WWW '21]{Proceedings of the Web Conference 2021}{April 19--23, 2021}{Ljubljana, Slovenia}
\acmBooktitle{Proceedings of the Web Conference 2021 (WWW '21), April 19--23, 2021, Ljubljana, Slovenia}
\acmPrice{}
\acmDOI{10.1145/3442381.3449824}
\acmISBN{978-1-4503-8312-7/21/04}

\begin{document}

\title{Detecting and Quantifying Wash Trading on Decentralized Cryptocurrency Exchanges}


\author{Friedhelm Victor}
\affiliation{%
  \institution{Technische Universit{\"a}t Berlin}
  \streetaddress{Straße des 17. Juni 135, 10623 Berlin}
  \city{Berlin}
  \country{Germany}}
\email{friedhelm.victor@tu-berlin.de}

\author{Andrea Marie Weintraud}
\affiliation{%
  \institution{Technische Universit{\"a}t Berlin}
  \streetaddress{Straße des 17. Juni 135, 10623 Berlin}
  \city{Berlin}
  \country{Germany}}
\email{weintraud@campus.tu-berlin.de}

\begin{abstract}
  Cryptoassets such as cryptocurrencies and tokens are increasingly traded on decentralized exchanges. The advantage for users is that the funds are not in custody of a centralized external entity. However, these exchanges are prone to manipulative behavior. In this paper, we illustrate how wash trading activity can be identified on two of the first popular limit order book-based decentralized exchanges on the Ethereum blockchain, IDEX and EtherDelta. We identify a lower bound of accounts and trading structures that meet the legal definitions of wash trading, discovering that they are responsible for a wash trading volume in equivalent of 159 million U.S. Dollars. While self-trades and two-account structures are predominant, complex forms also occur. We quantify these activities, finding that on both exchanges, more than 30\% of all traded tokens have been subject to wash trading activity. On EtherDelta, 10\% of the tokens have almost exclusively been wash traded. All data is made available for future research. Our findings underpin the need for countermeasures that are applicable in decentralized systems.

\end{abstract}



\maketitle

\section{Introduction}
Despite several cycles of boom and bust, cryptocurrencies and other virtual assets in the form of blockchain-based tokens appear to remain popular. Founders of blockchain startups frequently choose to issue a token, e.g. to represent shares or voting rights in the startup.
blockchain-based token systems became popular in 2017, and have evolved in recent years. In fact, some new developments even see heightened attention:
A new group of services under the umbrella term \textit{Decentralized Finance} (DeFi) are establishing themselves as a key use case for distributed ledger technologies (DLTs) in general. The aim is to design financial products, that are partly based on services from the traditional financial world yet completely new in other areas.
The appeal lies in offering these financial products in a decentralized way, excluding middlemen as far as possible.

While the field now encompasses a multitude of services such as lending protocols, derivatives and insurances, so-called decentralized exchanges (DEX) were among the early drivers of the ecosystem.
They allow for the exchange of virtual assets without having to rely on externally-controlled services such as centralized exchanges, many of which have been the victims of theft in recent years~\cite{mccorry2018preventing}.
As the number of different DeFi services increases, so does the complexity and vulnerability of these systems, as demonstrated by the recent example of large so-called flash loan-based trades. Within a single transaction, a large cryptocurrency loan is taken and repaid while executing an attack vector against multiple services~\cite{qin2020attacking}.
It is therefore necessary to gain a better understanding of how services in the DeFi ecosystem are being used. This can help to uncover manipulative behavior, so that these systems can be improved in the future.

In this work, we therefore focus on two of the first and most popular decentralized exchanges on the Ethereum platform, where we study the trading activity between user accounts. In particular, we identify a lower bound of suspicious trading behavior that closely follows the definition of wash trading, a type of market manipulation that is well known in the context of traditional financial instruments, and illegal in most countries. 
To perform wash trading, several users can collude and trade only amongst themselves. Thereby, they give the impression that they are buying and selling, but in reality they are not changing their own positions or taking any real market risk. These activities inevitably lead to increased (fake) trading volume, a metric that is observed among many traders and may influence a trader's sentiment about a given virtual asset. The same effect can also be achieved with a single user operating multiple accounts. Since account creation on the Ethereum blockchain is virtually cost-free, and does not require providing identity information, this scenario is much more likely.

Although wash trading in cryptocurrencies has been studied on centralized exchanges~\cite{cong2019crypto}, where trading happens off-chain, previous analysis is limited to exchange-reported trades, for which account level information is unavailable.
In addition, it is very likely that these reported trades have never actually taken place, and the exchange operators are the main suspects. This is in contrast to trades on decentralized exchanges, where each trade carries a cost and is stored on the ledger along with the involved accounts.

\noindent To the best of our knowledge, no systematic analysis of wash trading activity exists for decentralized exchanges.

\noindent \textbf{Contributions in this paper}:
\begin{itemize}
  \item We present the first systematic analysis of wash trading behavior on the decentralized exchanges IDEX and Etherdelta, demonstrating the issue and the need for countermeasures.
  \item We empirically identify wash trade accounts and trade structures and discover that the majority of structures consist of one or two accounts, while complex ones also exist.
  \item We quantify a lower bound of the financial extent of the activities, finding that on both exchanges, a total of 159 million U.S. Dollars has been wash traded, that at least 30\% of all tokens on both exchanges experienced wash trading activity, and that on EtherDelta, 10\% of the traded tokens have almost exclusively been wash traded.
\end{itemize}

\section{Background}
With the inception of Bitcoin~\cite{nakamotobitcoin}, digital currencies have gained in popularity. Users can manage their currency within an account, also called a wallet, which consists of a public and private key pair. To send a transaction, users have to specify a target account address, which is derived from a public key, and sign it with their own private key. Today, users are not limited to transferring the native cryptocurrency of a particular blockchain. The emergence of blockchains featuring smart contracts has paved the way for the issuance of a wide variety of assets that are typically represented in the form of tokens. Ethereum~\cite{wood2014ethereum} is a prime example for this, where thousands of such tokens have been issued. It is also a platform that has seen widespread use of decentralized exchanges, where tokens can be traded against Ether (ETH) and other tokens, thus offering an alternative to external, centralized exchanges.

\subsection{Centralized and Decentralized Exchanges}
A centralized exchange (CEX), sometimes also called a custodial exchange, keeps a user's assets in their collective exchange wallets.
These are high-value targets of attacks, as evidenced by numerous cryptocurrency exchange hacks. In order to trade on a CEX, users send their assets to an exchange-owned deposit wallet specifically created for a particular user. This wallet then forwards the assets to their main wallet and registers the received funds to be used for trading. All trading happens outside of the blockchain, until a user wants to withdraw their assets, at which point the exchange sends them back to the user wallet.

In contrast, a decentralized exchange (DEX) is typically implemented as a smart contract, which can allow for the non-custodial trading of cryptoassets.
Users send their assets to the smart contract, can interact with it for trading, and withdraw again. In the past years, multiple types of DEX have been proposed, some of which are actively used.

\subsection{Types of Decentralized Exchanges}
The two main DEX variants are based on limit order books (LOB) or automated market makers (AMM). The following is a brief overview. For more details, we refer the reader to Lin et al.~\cite{lin2019deconstructing}
\subsubsection{Limit Order Book-based DEX}
In a LOB-based DEX, users trade with each other via an order book. As makers, they submit orders to the order book as an offer to buy or sell an asset at a certain price and/or volume. When filling such an order, either explicitly or automatically, the users act as takers. The order books can be managed on-chain or off-chain, but the actual settlement of the trades typically happens on-chain. Two popular examples of such LOB-based DEX are EtherDelta and IDEX, that both operate on the Ethereum blockchain. They both manage their order books off-chain, which requires an external service that users interact with. Nevertheless, they can withdraw their assets at any time by interacting with the smart contract directly. In the case of EtherDelta, users send trade instructions to the smart contract itself, whereas IDEX uses a separate account to trigger trade instructions at the smart contract. They both charge trading fees, which amount to $0.3$\% of the traded amounts, where IDEX splits the fees between maker and taker, and EtherDelta charges only the taker. In addition, the transaction fees need to be covered by the traders, effectively increasing the total trading fees to more than $0.3$\%.

\subsubsection{Automated Market Maker DEX}
AMM DEX such as Uniswap and Kyber differ significantly from an LOB DEX. Users do not trade peer-to-peer but against a liquidity pool or reserve. These are implemented using smart contracts. On Uniswap, each reserve or contract holds funds of a token and ETH. Uniswap features a pricing mechanism called a "constant product market maker" formula. Prices are determined by the share of token and ETH funds in a reserve in relation to an invariant of total liquidity, which remains stable during trading. Liquidity providers fill the pools and must always contribute equal amounts of a token and ETH~\cite{Uniswap-v1-whitepaper}.

Although AMM DEX have recently gained in popularity, we focus on LOB-based DEX in this work, as they have a longer history of over 3 years.

\subsection{Defining Wash Trades}

Wash trading has been prohibited in the U.S. by the Commodity Exchange Act in 1936 \cite{cftc-cea} \cite{cea-full}. The Commodity Futures Trading Commission (CFTC) defines it as "Entering into, or purporting to enter into, transactions to give the appearance that purchases and sales have been made, without incurring market risk or changing the trader's market position" \cite{cftc-wash-trading}. It is also referred to as Round Trip Trading. Actors colluding in or arranging transactions, such that they do not incur market risk, are thereby deemed wash trades. The definition also indicates that actors executing a series of transactions among themselves, after which they end up at the same market position that they had initially, can be considered wash trades.
In 2013, the CME Group has issued a Market Regulation Advisory Notice. It provides some guidance as to what may constitute a wash trade. It notes two important criteria: a) the intent of actors to execute fictitious trades that are not subject to market risk, and b) "a wash result - meaning the purchase and sale of the same instrument at the same price, or a similar price, for accounts with the same beneficial ownership or for accounts with common beneficial ownership" \cite[p. 5]{cme-letter-wash-trading}.

\section{Related Work}
Academic studies that are related to our work mostly concern wash trading in traditional financial markets and other types of market manipulation in cryptocurrencies. While these will be addressed in detail in the following, other related works concern the graph-based study of token transfer networks~\cite{chen2020traveling,somin2018network,victor2019measuring} and the identification of users with multiple accounts in Ethereum~\cite{victor2020address}.

\subsection{Cryptoasset Price Manipulation}
The increasing attention for cryptocurrencies has come with a range of manipulative activities.
Several papers have studied the phenomenon of cryptocurrency pump and dump schemes~\cite{victor2019cryptocurrency, xu2019anatomy, pump-dump-2019-li,pump-dump-2018-kamps} and manipulation of Bitcoin prices~\cite{price-manipulation-bitcoin-2020,price-manipulation-bitcoin-2018}, both on centralized exchanges.
Suspicious trader networks and manipulative transaction patterns have been identified on the Mt. Gox exchange \cite{market-manipulation-bitcoin-2019}.
Much less work exists on manipulation on DEX, though substantial evidence for front running attacks was found \cite{daian2019flash}.

\subsection{Wash Trade Detection}
Before academic work has focused on the detection of wash trading, a related phenomenon called collusive cliques has been studied using graph clustering and nearest neighbor algorithms~\cite{collusion-graph-2005}. Other methods that have been used include spectral clustering~\cite{franke-2008-spectral-clustering}, Hidden Markov Models~\cite{cao-2010-abnormal-coupled-sequences} and correlation statistics on aggregated order volume time series \cite{collusive-cliques-2011}.

Cao et al. were among the first to explicitly study the detection of wash trades in financial markets \cite{cao2014detecting}. They point out that previous work had focused on collusive and correlated trading behavior rather than the more specific pattern of wash trading.
Their detection algorithm identifies subsets of a given set of trades that lead to no position change of all involved traders. From a topological perspective, these trades form a closed cycle. To identify these, the algorithm checks the power set of a set of trades, resulting in a runtime of $\mathcal{O}(2^n)$.
Cao et al. extended their work in 2016 for order book data instead of trade data \cite{cao2016detecting}.

\subsection{Cryptocurrency Wash Trading}
Wash trading in cryptocurrencies is receiving increasing attention.
Websites have published articles on a startup faking volume for exchanges \cite{coindesk2019fakevolume} and exchanges have promised to come clean on wash trading \cite{website-okex-wash-trading}.
Recently, the first academic study on wash trading on CEX has been proposed by Cong et al. \cite{cong2019crypto}. The authors study major cryptocurrencies, finding that 26 unregulated exchanges fake up to 70\% of their total trading volume. They analyze first-significant-digit distributions of trade sizes, and size clustering and power laws of trade size distributions. In Section~\ref{high_level_insights}, we briefly show that trade size distributions are of little use for detecting wash trading on DEX.
Cong et al., as well as some previous technical reports, have not analyzed wash trading on an account-level basis, but rather used statistical indicators to find evidence for fake volume.
Also, it is likely that such fake volumes or wash trades are produced by the exchanges themselves, to boost their trading volumes and thereby rankings. However, individual traders can also perform wash trading, but previous research on CEX was unable to perform such analysis due to a lack of data. To this end, we analyze wash trading on DEX on an account-level basis, following legal definitions.

\section{Datasets and Preparation}
In this section, we detail what data we have collected and how it was preprocessed. We also provide information on how we model token trade graphs, and provide high level insights about the datasets.

We collected all Ethereum blocks, transactions and events related to the EtherDelta and IDEX exchanges, corresponding to a time frame between \nIDEXandEDStartDate and \nIDEXandEDEndDate.
Given the transactions and events, along with the smart contract addresses of IDEX and EtherDelta, information about the executed trades can be obtained.

To extract trades that have been executed on IDEX, we have parsed all successful transactions targeted at the IDEX smart contract\footnote{IDEX smart contract address: 0x2a0c0dbecc7e4d658f48e01e3fa353f44050c208}, which call the function \texttt{trade}, identified by the shortened Keccack256 hash \texttt{0xef343588}. The parameters of this function call include information on the participating accounts, which token has been bought and sold, traded amounts and fees.

Similarly, EtherDelta trades can be extracted by parsing the trade events triggered by the EtherDelta smart contract\footnote{EtherDelta smart contract address: 0x8d12a197cb00d4747a1fe03395095ce2a5cc6819}.
Every successful EtherDelta trade emits a \texttt{Trade} event, identified by the event topic Keccack256 hash starting with \texttt{0x6effdda7}, which is triggered by the smart contract when a corresponding transaction has been executed against it. The event also contains information about the participating accounts, exchanged tokens and amounts.

The trade timestamps are obtained from the block the transaction or event was found in.
To assess the trade volume in U.S. Dollars, we obtained daily exchange rates of Ether to Dollar from Etherscan. All data is available on Zenodo\footnote{\url{https://zenodo.org/record/4540223}}, and the code is on GitHub.\footnote{\url{https://github.com/friedhelmvictor/lob-dex-wash-trading-paper}}

\subsection{Data Preprocessing}
We have preprocessed the data as follows:
As the internal trade amount representation is based on very large integer amounts for precision, we first converted them to floating point values for easier handling. Secondly, the prices of the involved assets are computed based on exchanged amounts, and U.S. Dollar values are joined. We then removed trades containing missing data and possibly unsuccessful IDEX transactions as indicated by the status field which has been introduced with the Ethereum Byzantium fork. We also removed trades in which two tokens were exchanged, leaving trades between Ether and some token. We do so in order to be able to study wash trading of individual tokens traded against a common base currency. These steps remove only a very small fraction of the trades (IDEX: \nIDEXTradesRemoved trades, EtherDelta: \nEtherDeltaTradesRemoved trades). An overview of the resulting trade datasets is depicted in Table~\ref{table:prep-data-summary}.

\begin{table}[!h]
    \centering
    \caption{Overview of IDEX and EtherDelta datasets.}
    \label{table:prep-data-summary}
    \begin{tabular}{c c c }
        \hline
        \hline
        & IDEX & EtherDelta \\
        \hline
        Start date (UTC) & \nIDEXStartDate & \nEtherDeltaStartDate \\
        End date (UTC) & \nIDEXEndDate & \nEtherDeltaEndDate \\
        Number of trades & \nIDEXTradesPreprocessed & \nEtherDeltaTradesPreprocessed \\
        Number of traders & \nIDEXAccountsPreprocessed & \nEtherDeltaAccountsPreprocessed \\
        Number of tokens & \nIDEXTokensPreprocessed & \nEtherDeltaTokensPreprocessed \\
        \hline
        \hline
    \end{tabular}
\end{table}

\begin{figure}[ht]
    \centering
    \begin{subfigure}[b]{0.49\linewidth}
        \centering
        \includegraphics[width=\linewidth]{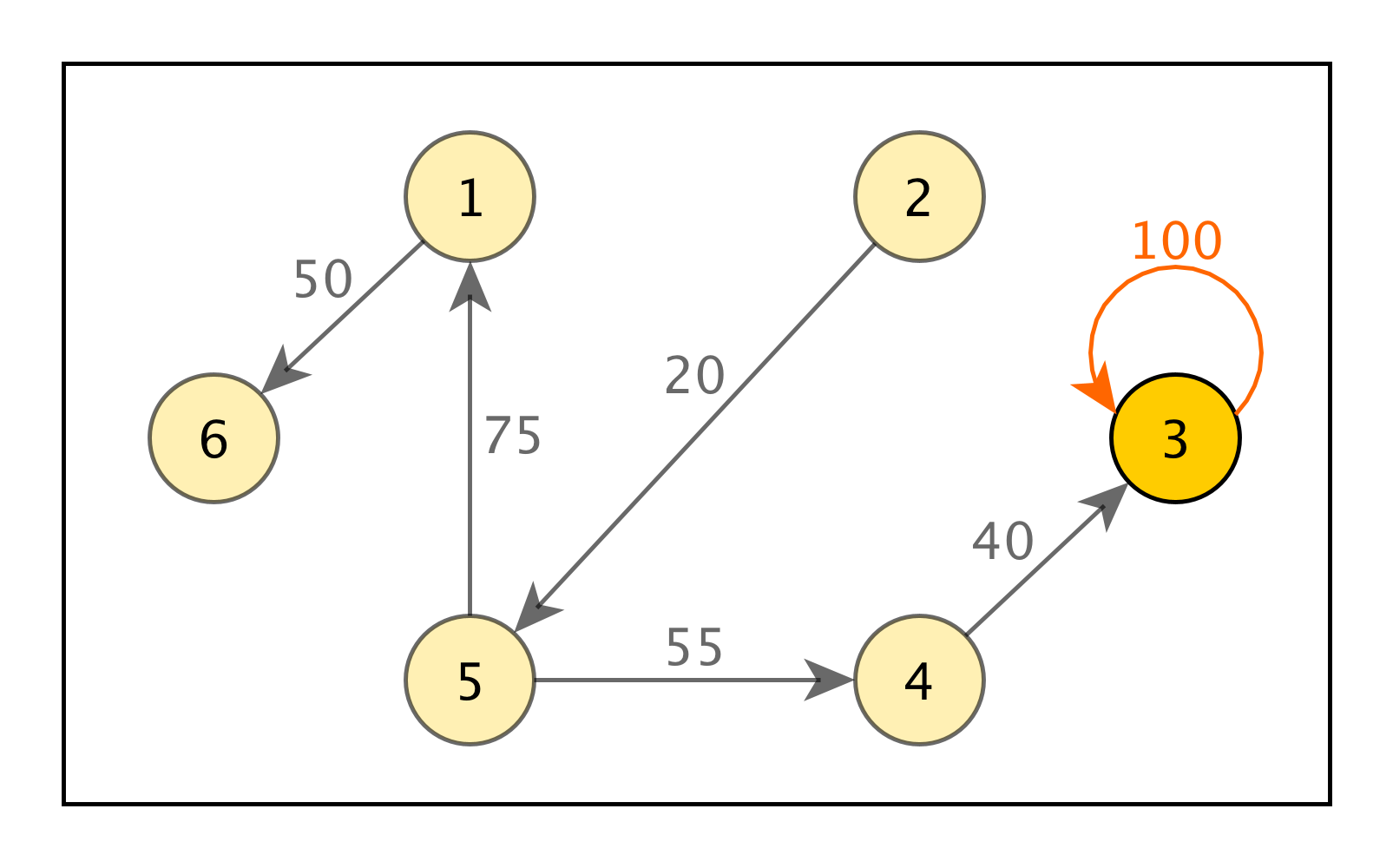}
        \caption{A Loop}
        \label{subfig:self-trade}
    \end{subfigure}
    \begin{subfigure}[b]{0.49\linewidth}
        \centering
        \includegraphics[width=\linewidth]{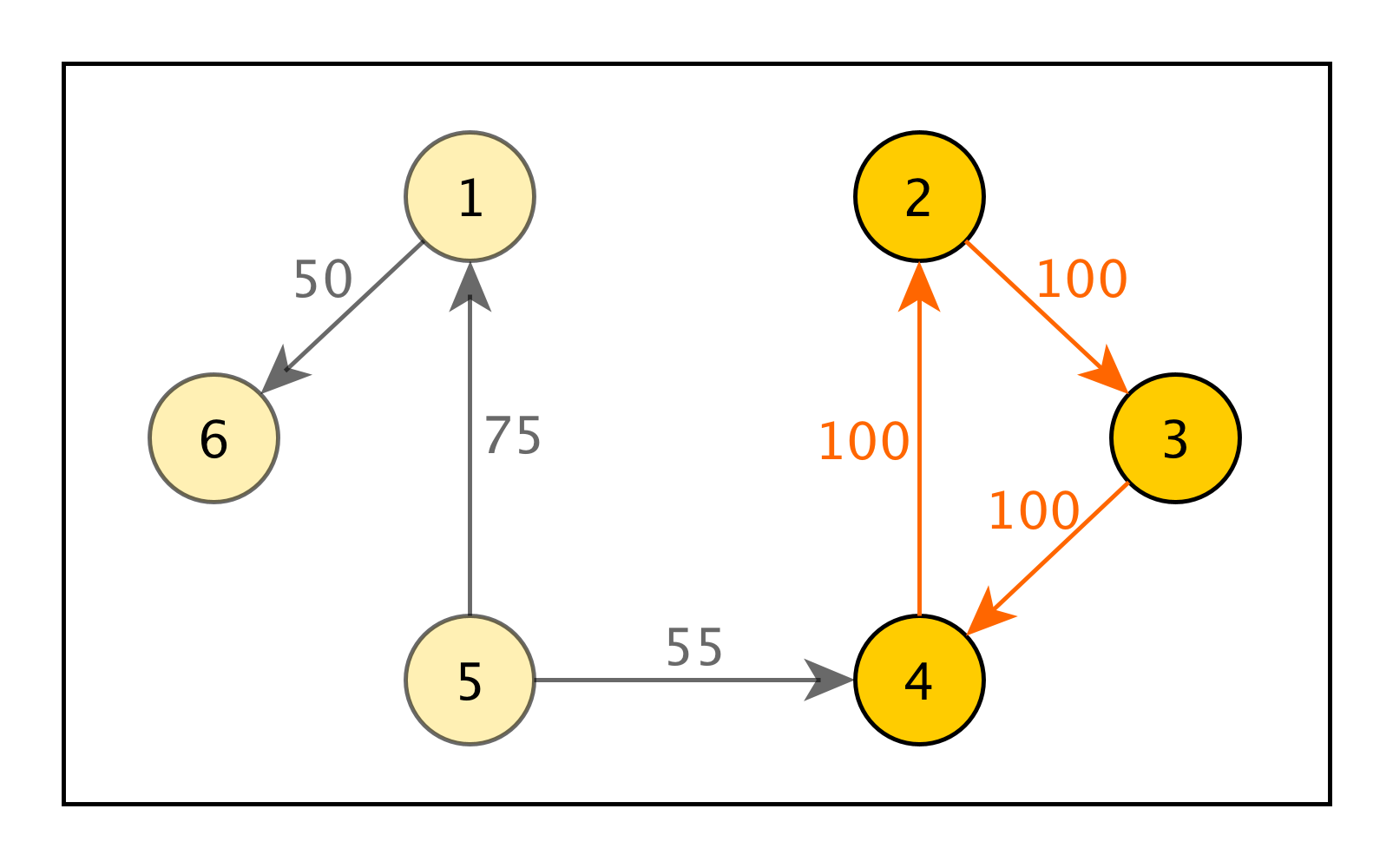}
        \caption{A Cycle}
        \label{subfig:wash1}
    \end{subfigure}
    \begin{subfigure}[b]{0.49\linewidth}
        \centering
        \includegraphics[width=\linewidth]{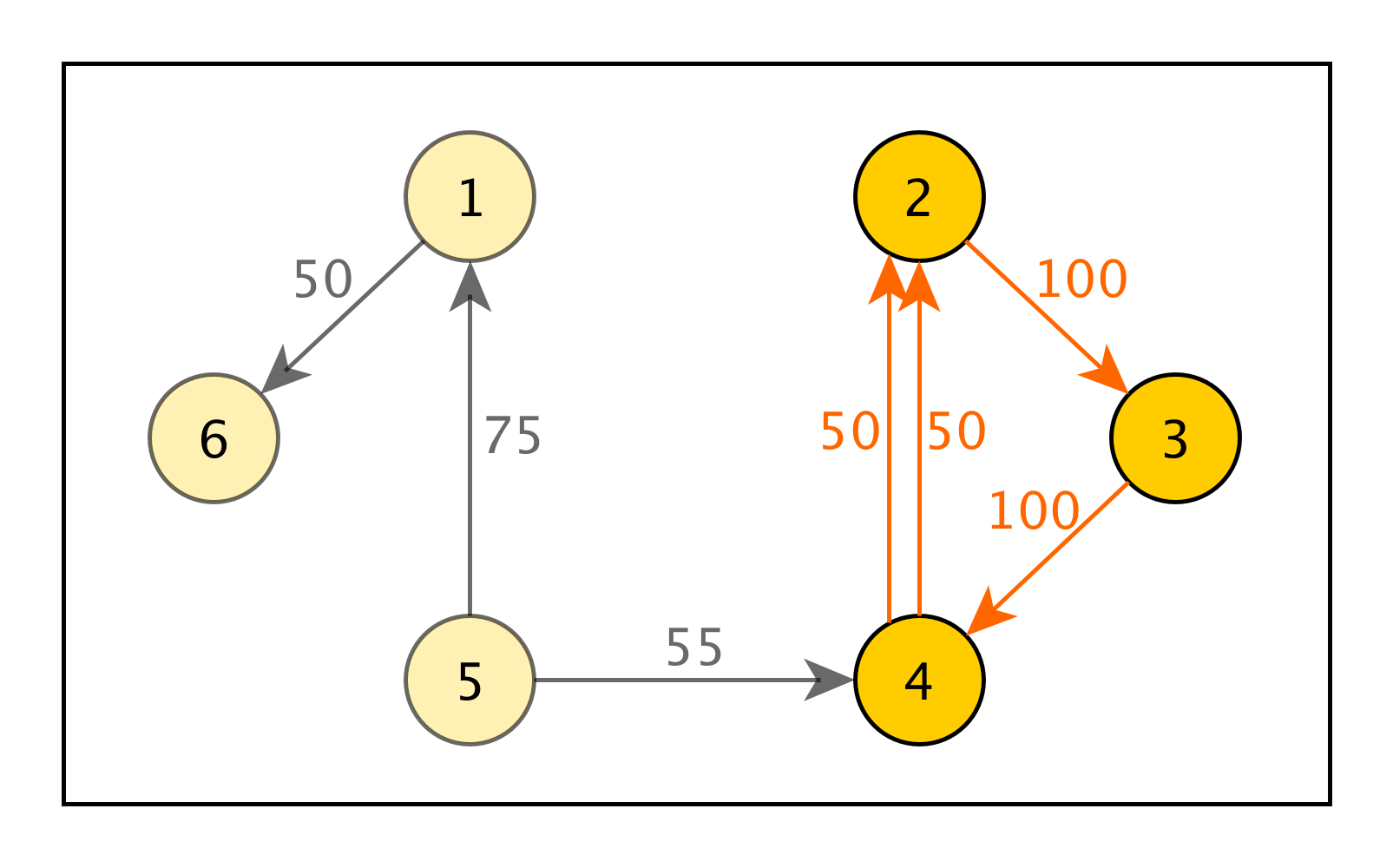}
        \caption{A Cycle With Parallel Edges}
        \label{subfig:wash2}
    \end{subfigure}
    \begin{subfigure}[b]{0.49\linewidth}
        \centering
        \includegraphics[width=\linewidth]{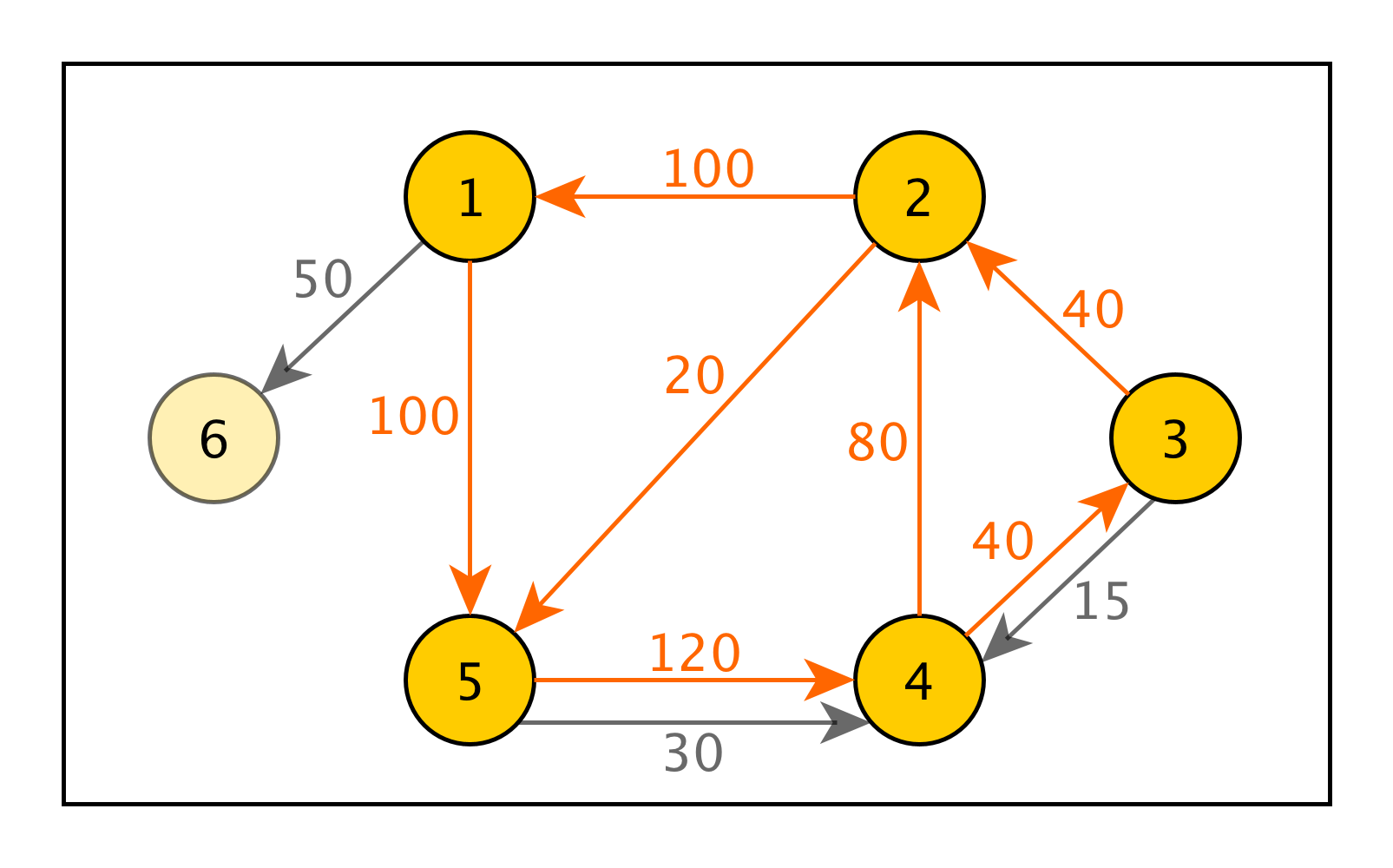}
        \caption{A Cycle with Sub-Cycles}
        \label{subfig:wash3}
    \end{subfigure}
    \caption{Four examples of directed token trade multigraphs. Red edges would conform to wash trading activity.}
    \label{fig:token-graph-wash-trade}
\end{figure}

\subsection{Modeling Token Trade Graphs}
Each trade in the data contains two participating accounts, and a token amount that was exchanged for a certain Ether amount. Therefore, we can model all trades of a certain type of token as a token trade graph. For the remaining chapters, we define a token trade graph $G(V,E)$ to be a directed multigraph, where $V$ is the set of trading account addresses, and $E$ is the set of trades. The direction of each edge is given by the token flow --- from the token seller to the token buyer.
Note that this means each trade is represented as only one directed edge, even though both vertices send and receive assets.
Finally, each edge also has a weight, which corresponds to the traded token volume.

This setup allows us to observe cycles with matching volumes in a token trade graphs, which is fundamental to the idea of wash trading.
To illustrate, Figure~\ref{fig:token-graph-wash-trade} showcases a few potential scenarios of wash trading activity within a token trade graph. Each vertex is one trading account, trades belonging to the wash trade set are colored red, and vertices belonging to the set of wash trading actors are colored in a strong yellow. The rest of the graph is slightly faded. Figure \ref{subfig:self-trade} shows a loop, i.e. a self-trade. This means a single account executed a trade with itself, which can also be observed in our data. In Figure \ref{subfig:wash1}, three actors trade the same amount in a cycle, slightly modified in Figure \ref{subfig:wash2}, where the trades between trader $4$ and $2$ are split into two trades. Finally, Figure \ref{subfig:wash3} shows a more complicated scenario. There is a cycle along vertices $1$ to $5$, but also three partly overlapping sub-cycles $\{1,5,4,2\}$, $\{2,5,4\}$ and $\{2,5,4,3\}$. Summing the trades, each trader buys and sells exactly 120 tokens, which leads to no change in their market positions. Also, note that not all trades even among the wash trading actors belong to the wash result, e.g. 30 token from trader $5$ to $4$. Indeed, actors may try to hide their illegitimate activities among other, legitimate trades.


\begin{figure}[ht]
    \centering
      \includegraphics[width=\linewidth]{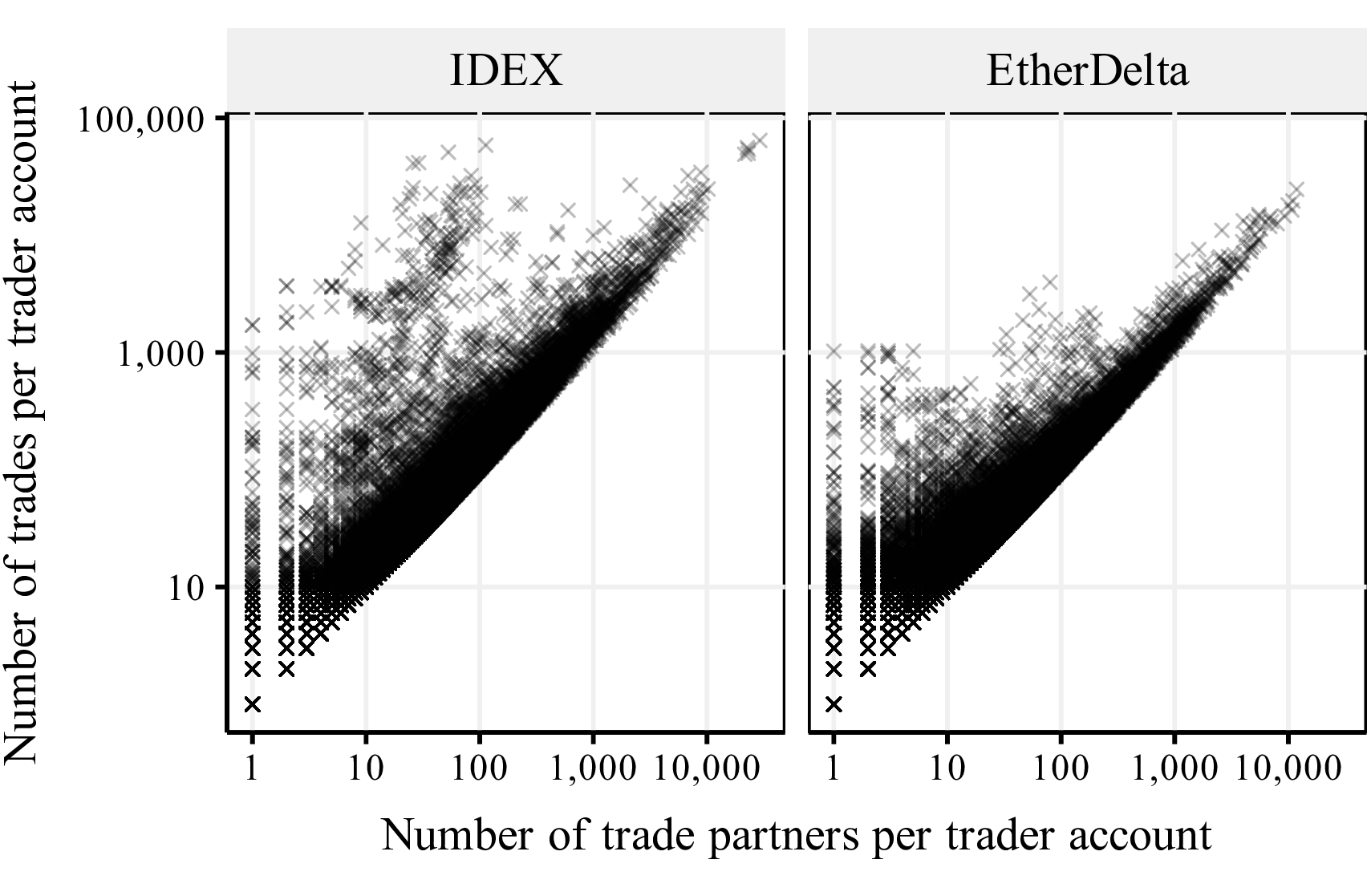}
    \caption{Each cross represents one trader account, positioned by number of trade partners and number of trades particpated in. Surprisingly, some traders perform a large number of trades with only a few other accounts.}
    \label{fig:trade_partners_trades}
\end{figure}

\begin{figure}[hb]
    \centering
      \includegraphics[width=\linewidth]{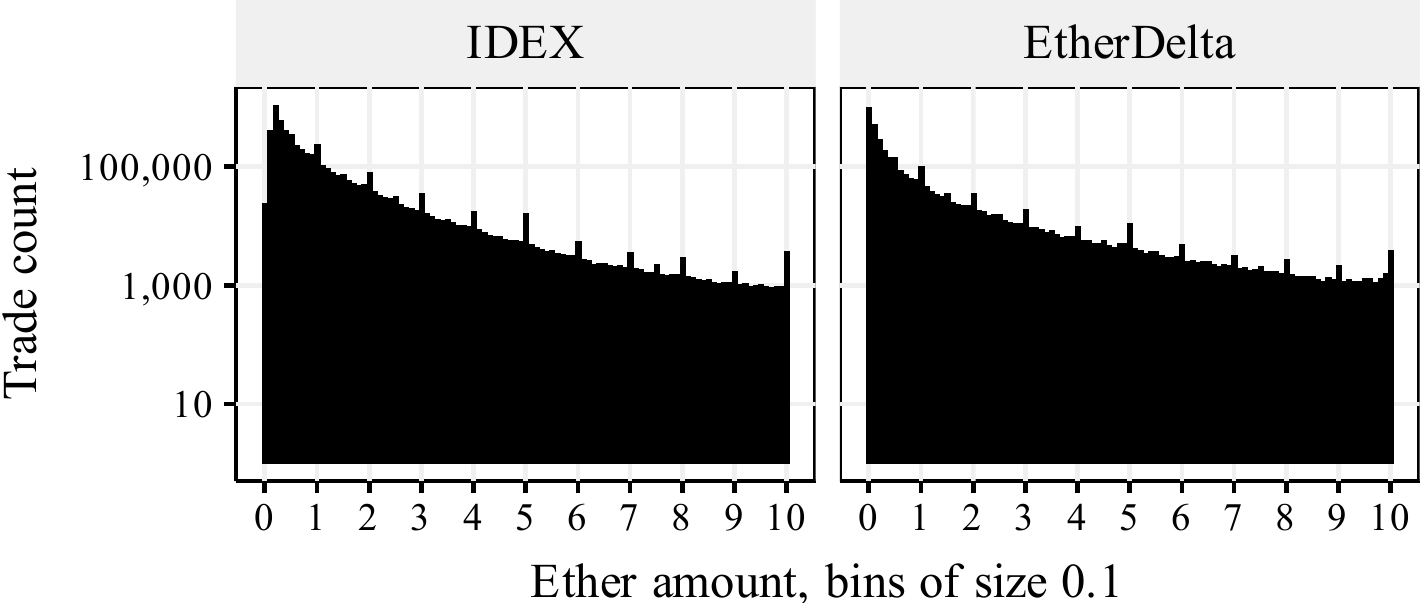}
    \caption{Trade size distributions in Ether amounts. Higher amounts are traded less frequently, and local peaks exist at round numbers, indicating natural trading behavior.}
    \label{fig:trade-size-dist}
\end{figure}

\subsection{High Level Insights} \label{high_level_insights}

We briefly explore the datasets to illustrate their structure.
Particularly interesting is the relationship between the number of trade partners and the number of executed trades, which is visualized in Figure~\ref{fig:trade_partners_trades}. Each cross is one trading account, positioned by the number of trades it was part of, and how many other trading accounts it has interacted with. On the left side of each plot, there exist accounts that have performed many trades with only a few other accounts.
The ratio of number of trades per trade partner over all trader accounts has its median at just 1, and is only 3.25 at the 99th percentile. This makes sense intuitively, as traders usually do not know who they are trading with when issuing a trade order. The DEX web interfaces that list current buy and sell orders do not indicate account information. Therefore, accounts that have vastly more trades than trade partners might be suspicious.

Secondly, we can explore traded Ether amounts. Related works use the indicator of an unnatural trade size distribution (such as a uniform distribution) as a signal for the existence of wash trading on centralized exchanges. Figure ~\ref{fig:trade-size-dist} shows the trade size distribution for both DEX analyzed in this paper. Higher amounts are traded less frequently, and there exist local peaks at round numbers such as 5 Ether.
We conclude that these distributions appear to conform to natural trading behavior, and are likely unsuitable as an indicator for wash trading on DEX.

\begin{figure*}[!htbp]
    \centering
    \includegraphics[width=\textwidth]{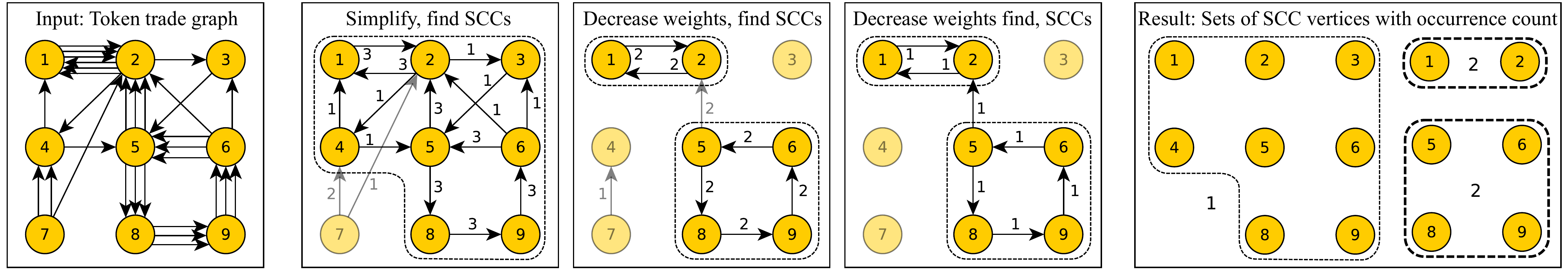}
    \caption{After graph simplification, SCCs are iteratively counted, and weights decreased until no edges remain (Algorithm~\ref{algo:iterative-scc}).}
    \label{fig:SCC-layers}
\end{figure*}

\section{Wash Trade Detection}

In order to detect wash trades, we follow legal definitions and criteria of wash trading by the CFTC and CME.
Our aim is therefore to identify sets of trades between collusive trading accounts that lead to no change in the individual position of each participating trader.
In other words, for each account within a set of trading accounts, the total amount of purchased assets equals the total amount of sold assets, such that the involved traders essentially hold the same position they had initially. In this section, we present a volume matching algorithm that identifies such trades, which needs to be supplied with a set of trades and a time window parameter.
Due to the very high number of trades in our datasets, it is not feasible to check the power set of the trade sets as proposed with previous wash trade detection methods, as this results in exponential time complexity.
For this reason, we follow a two-step process:
\begin{enumerate}
    \item \emph{Account Candidate Set Generation}: We determine candidate sets of potentially collusive traders via an iterative counting of strongly connected components (SCCs) in each token trade graph.
    \item \emph{Trade Volume Matching}: For time windows of trades within frequently occurring SCCs, we determine if there exists a trade subset that leads to no position change for each trader.
\end{enumerate}

\subsection{Candidate Set Generation with SCCs}\label{sec:candidate-scc}
We saw in Figure~\ref{fig:token-graph-wash-trade}, that wash trade scenarios contain at least one cycle. However, an approach of detecting all cycles would not be able to capture the scenario in Figure~\ref{subfig:wash3}, where the wash trading structure consists of a cycle with sub-cycles. Therefore, we propose to use the concept of strongly connected components to identify maximal subsets of vertices that are connected via cycles.

A directed graph $G$ is strongly connected if there exists a directed path from every vertex $v$ to every other vertex $u$ in $V$. If $G$ as a whole is not strongly connected, it can consist of strongly connected components (SCCs). An SCC is defined as a maximal subset of vertices $C \subseteq V$ such that there exists a directed path from every vertex $u$ to $v$ and from $v$ to $u$, where $u, v \in C$~\cite{algorithms-online}. An SCC contains one or multiple cycles, which can be thought of as round trip trades.
As such cycles can appear coincidentally, we are particularly interested in traders that are repeatedly part of such SCCs. Therefore, for each token trade graph, we perform an iterative counting of SCCs as illustrated in Algorithm~\ref{algo:iterative-scc} and Figure~\ref{fig:SCC-layers}.

\begin{algorithm}[ht]
\SetAlgoLined
\DontPrintSemicolon
\KwData{Directed token trade multigraph $G(V,E)$}
\KwResult{Mapping $M$ of sets of vertices $V$ to counts}
 $G \leftarrow$ simplify(G, edgeweight $=$ edge count)\;
 $M \leftarrow$ Create empty key value map, default value $0$\;
 \While{$|E| > 0$} {
   $SCCList \leftarrow$ computeSCCs(G)\;
   \ForEach{$SCC \in SCCList$}{
     $SCC_V \leftarrow$ getVertexSet($SCC$)\;
     $M.set(SCC_V) \leftarrow M.get(SCC_V)+1$\;
   }
   \ForEach{$e \in E$}{
     $weight(e) \leftarrow weight(e)-1$\;
     \If{$weight(e) = 0$}{
       $E \leftarrow E \setminus \{e\}$\;
     }
   }
 }
 \Return(M)\;
\caption{Iterative SCC Counting}
\label{algo:iterative-scc}
\end{algorithm}

First, a directed token trade multigraph is simplified, turning the number of multi edges between two vertices into an edge weight. We then perform multiple iterations of determining all SCCs in the graph, counting how often they have been seen over the iterations. After each iteration, we decrease all edge weights by $1$, and remove edges that would end up with an edge weight of $0$. When no edges remain, the process is complete. The result are sets of vertices of SCCs with counts indicating how often they have occurred.

\begin{figure}[h]
    \centering
    \includegraphics[width=\linewidth]{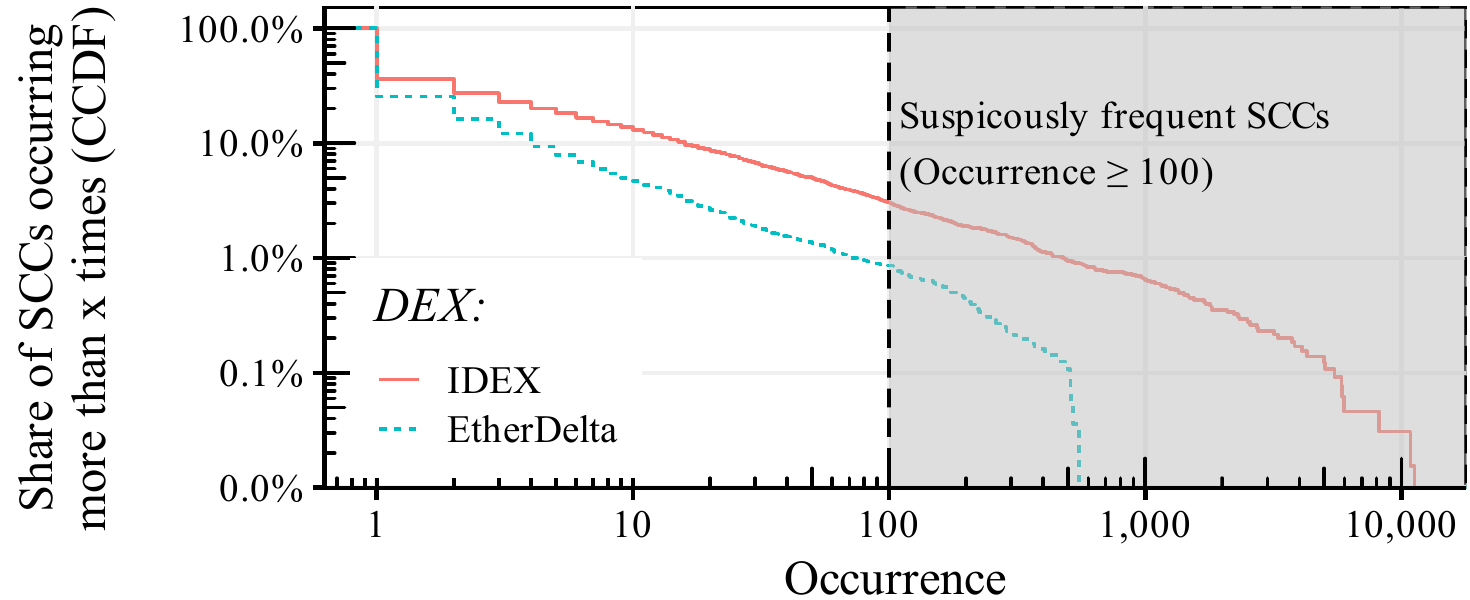}
    \caption{CCDF of identified SCCs. A small fraction of SCCs appear at least 100 times, indicating at least 100 round trip trades. We choose these as our candidate set.}
    \label{fig:SCC-threshold}
\end{figure}

After performing Algorithm~\ref{algo:iterative-scc} on all token trade graphs of each DEX, we can visualize the complementary cumulative distribution function (CCDF) of SCC counts in Figure~\ref{fig:SCC-threshold}. We select those SCCs as our candidate set which occur at least 100 times. On IDEX, this corresponds to the top $3\%$, on EtherDelta the top $1\%$.

\subsection{Volume Matching}

After identifying candidate sets of SCCs consisting of trading accounts that repeatedly trade in circular ways, we look at the trades between these accounts, for individual tokens. The goal is to find subsets of trades with matching volumes that lead to no overall position change for the individual accounts. In this section, we therefore formalize position changes based on multiple trades, adapted from Cao et al.~\cite{cao2014detecting, cao2016detecting} for the context of crypto tokens. Finally, we describe our volume matching algorithm of linear time complexity.

\subsubsection{Formalizing trader position changes for wash trade detection.}
In order to detect trades that lead to no individual position change, we need to sum up trades in such a way that the traded volumes are summed up per trading account involved in these transactions. A volume can be bought or sold, so the position an individual account holds can increase or decrease.

We can formalize a trade $T$ as in Equation (\ref{eq:signed-trade}), where trader $a_j$ sells volume $V$ to trader $a_i$. We denote trader positions by their account addresses $a$, since one real-world trader may control several account addresses. The signs indicate the selling or buying position.
Since the algorithm that is based on this formalization is run for sets of trades within a particular token, we do not need to indicate the token here.

\begin{equation}
    \begin{split}
        T = (+a_i, -a_j)_V,
    \end{split}
    \label{eq:signed-trade}
\end{equation}

\noindent Two (or more) trades are summed up and lead to signed trader positions as in Equation (\ref{eq:sum-trades-position}). Buyer and seller of each trade are listed separately, and the sign for indicating selling and buying positions ($\pm$) is transferred to the volume $V$. The volumes sold and bought by each trader are summed up, leading to individual positions $p_i$. The result is a set of accounts $a_i$ that each hold a relative position $p_i$, which can be positive or negative. We refer to the overall position of individual traders involved in a set of trades $T_k$ as $P$.

\begin{equation}
    \begin{split}
        \sum_{k=1}^{2} T_k & = T_1 + T_2 \\
        & = (+a_i, -a_j)_{V_1} + (+a_j, -a_i)_{V_2} \\
        & = \{ [+a_i]_{V_1}, [-a_i]_{V_2}, [-a_j]_{V_1}, [+a_j]_{V_2} \} \\
        & = \{ [a_i]_{+V_1 -V_2}, [a_j]_{-V_1 +V_2} \} \\
        & = \{ [a_i]_{p_i}, [a_j]_{p_j} \} \\
        & = P
    \end{split}
    \label{eq:sum-trades-position}
\end{equation}

\noindent We formalize the detection of wash trades with Equation (\ref{eq:zero-sum}). For a given trade set $S_T$ the goal is to find a subset $S_W \subseteq S_T$ such that:
\begin{equation}
\begin{split}
    \sum S_W & = \sum T_k \in S_W \\
    & = \sum (+a_i, -a_j)_{V_k} \\
    & = \{ [a_i]_{p_i}, ..., [a_j]_{p_j} \} \\
    & = \{ [a_i]_{0 \pm v}, ..., [a_j]_{0 \pm v} \}
\end{split}
\label{eq:zero-sum}
\end{equation}

\noindent In other words, when summing up all trades in $S_W$, all traders $a_i$ involved in $S_W$ must hold a relative position of $p_i=0$, allowing for a small deviation $v$.
We define this deviation as a percentage (or margin) $m$ of the mean trading volume of $S_W$, e.g. $m=0.01$:

\begin{equation}
    v = m \cdot \frac{\sum_{T_k \in S_W} V_k}{|S_W|}
    \label{eq:margin}
\end{equation}

We need to choose such a margin for two reasons: In order to make a wash trade less obvious, traders might not buy and sell exactly the same volumes when wash trading. Secondly, each trade incurs a cost, which may decrease a trader's capital with every trade and therefore also be reflected in slightly decreasing wash trade volumes. We set this margin to be $1\%$.

\subsubsection{Wash Trade Detection Algorithm.}
Based on the definitions, we can now determine whether a given set of trades yields trader positions that are close to $0$.
The next challenge is to test various trade sets. Given a set of trades, the algorithm by Cao et al.~\cite{cao2016detecting} tests all possible subsets (i.e. the power set), trying to find configurations such that the signed sums of each trader equal zero.
However, this leads to an exponential time complexity in $\mathcal{O}(2^n)$, in dependence on the size of the given trade set $n$.
In Section \ref{sec:candidate-scc}, we have proposed to generate SCC candidate sets per token trade graph in order to reduce the trade sets to be analyzed. However, these SCCs can still execute thousands of trades in a particular token, so an exponential runtime is not feasible. We therefore propose to further split these trades using a time window $t$. Also, instead of checking all $2^n$ subsets of a window trade set, we propose to check $n$ subsets, iteratively removing the last trade in the set.

Algorithm~\ref{algo:vol-match} illustrates this concept for a set of trades $S_T$. The signed trader positions are calculated by summing up all given trades (Line 3). The algorithm then iterates through the trades. In each step, it checks whether each trader has a current position $p_i$ of at most a margin $m$ of the mean trade size.
If the current trade set $S_W$ meets this condition, a wash trade set is found. If not, the last trade is repeatedly removed from the set (Lines 9-10).

\begin{algorithm}[hb]
\SetAlgoLined
\DontPrintSemicolon
\KwData{Set of trades $S_T = \{ (+a_i, -a_j)_{V_1}, ..., (+a_x, -a_y)_{V_n} \}$, margin $m$}
\KwResult{Set of trades that constitute a wash result $S_W$}
$S_W \leftarrow S_T$ \tcp*{assume all trades are in wash result}
 $n \leftarrow |S_W|$\;
 $ P \leftarrow \sum S_W = \{ [a_i]_{p_i}, ..., [a_j]_{p_j} \} $ \tcp*{compute positions according to Equation (\ref{eq:sum-trades-position}) }
 \While{n > 1} {
    $v_{mean} \leftarrow \frac{\sum_{T_k \in S_W} V_k}{|S_W|} $ \tcp*{compute mean trade size}
    \tcp{if all trader positions are at most a margin of the mean trade size:}
    \If{$(p_i \leq m \cdot v_{mean}) \forall [a_i]_{p_i} \in P $} {
        \Return{$S_W$}\;
    }
    $n \leftarrow n-1$\;
    $S_W \leftarrow S_W[1, ..., n]$ \tcp*{remove last trade}
    $ P \leftarrow \sum S_W = \{ [a_i]_{p_i}, ..., [a_j]_{p_j} \} $ \tcp*{update positions}
 }
\caption{Trade Volume Matching}
\label{algo:vol-match}
\end{algorithm}

\begin{figure*}[t]
    \centering
    \includegraphics[width=\textwidth]{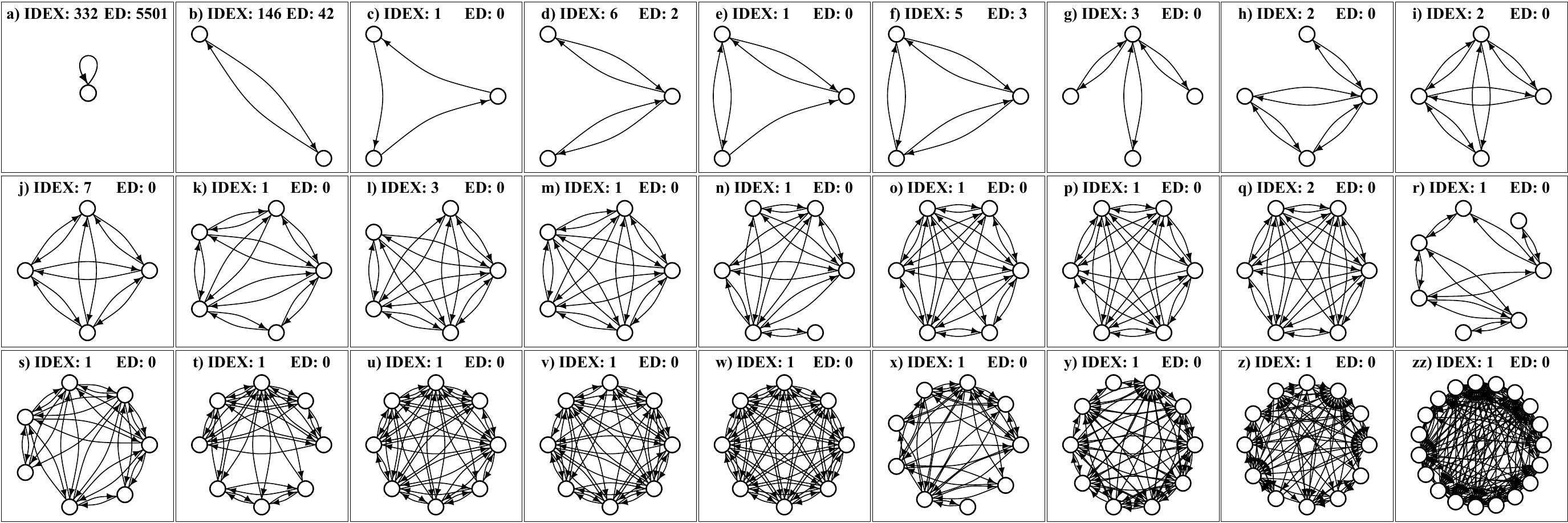}
    \caption{Identified wash trading structures with counts on both DEX. On EtherDelta (ED), the vast majority of wash trades are performed as self-trades of single accounts (5,501 instances). On IDEX, these are less frequent and two-account structures are also popular. More complex structures are rare and almost never appear on EtherDelta. It is noticeable that in almost all structures short sub-cycles of length 2 exist. Several cases -- b), f), j), m), q), w) -- are fully connected.}
    \label{fig:structures}
\end{figure*}

The individual positions are updated (Line 11) and checked again in the next iteration.
Finally, the algorithm terminates after removing the second to last trade (Line 4), as a wash trade set must contain at least 2 trades.
The complexity of the algorithm is in $\mathcal{O}(2n)$.

Wash trading actors may follow different temporal patterns and trade over different time periods until they reach what is defined as a wash result. Generally, we assume that such actors trade rather frequently in small time intervals so as to continuously inflate trading volumes that suggest an active market. Two matching wash trades with days in between them are less suitable for this purpose. However, in order to capture as many scenarios and wash results as possible, we propose to use three different time window sizes of one hour, one day and one week. Starting with the smallest interval, these are applied in three passes. In each pass, Algorithm~\ref{algo:vol-match} is applied on trade sets derived per candidate SCC, token and time window. Trades once labeled as wash trades are not checked again.

\section{Quantification}
In this section, we quantify the wash trading that has been identified by our detection method.
The following list summarizes the final parameters that form the basis for the remaining quantification of wash trading activity on IDEX and EtherDelta:
\begin{itemize}
  \item Threshold for candidate set of SCCs: $100$ occurrences across all traded token types.
  \item Margin $m$ at which we define that the trader position basically did not change: $1\%$ of the mean trade volumes within a set of analyzed trades.
  \item Time windows of trades in which we perform volume matching: $1$ hour, $1$ day and $1$ week.
\end{itemize}

\noindent We now study wash trading structures, what proportion of tokens is affected to what degree, at what point wash trading occurs within a token's trading lifecycle, and manipulated volume over time.

\subsection{Wash Trading Structures}
All trades of the candidate set SCCs were labeled with our volume matching algorithm. As a result, we can study the structures that consist of identified wash trades. Figure~\ref{fig:structures} illustrates the result for both IDEX and EtherDelta, where the counts indicate how many variants with different traders have been observed on each DEX. On EtherDelta, the vast majority of wash trades are performed as self-trades (a)), where an account is able to trade with itself, which could easily be prevented. Although also common, they do not appear to the same extent on IDEX. Second only to the self-trades, the structure consisting of two accounts (b)) is the next simplest, and at least on IDEX it is also quite common. More complex structures consisting of three or more accounts are mainly found on IDEX.
In almost every advanced structure, several short sub-cycles can be found.
In some instances the wash trade subgraphs are completely connected. The complex structures consisting of simpler sub-structures may indicate that wash trading is performed in simple cycles on a low-level, but that actors make an effort to hide these among multiple trading accounts. Branched structures with sub-cycles of length greater than three do not occur here.

\begin{figure}[ht]
    \centering
      \includegraphics[width=\linewidth]{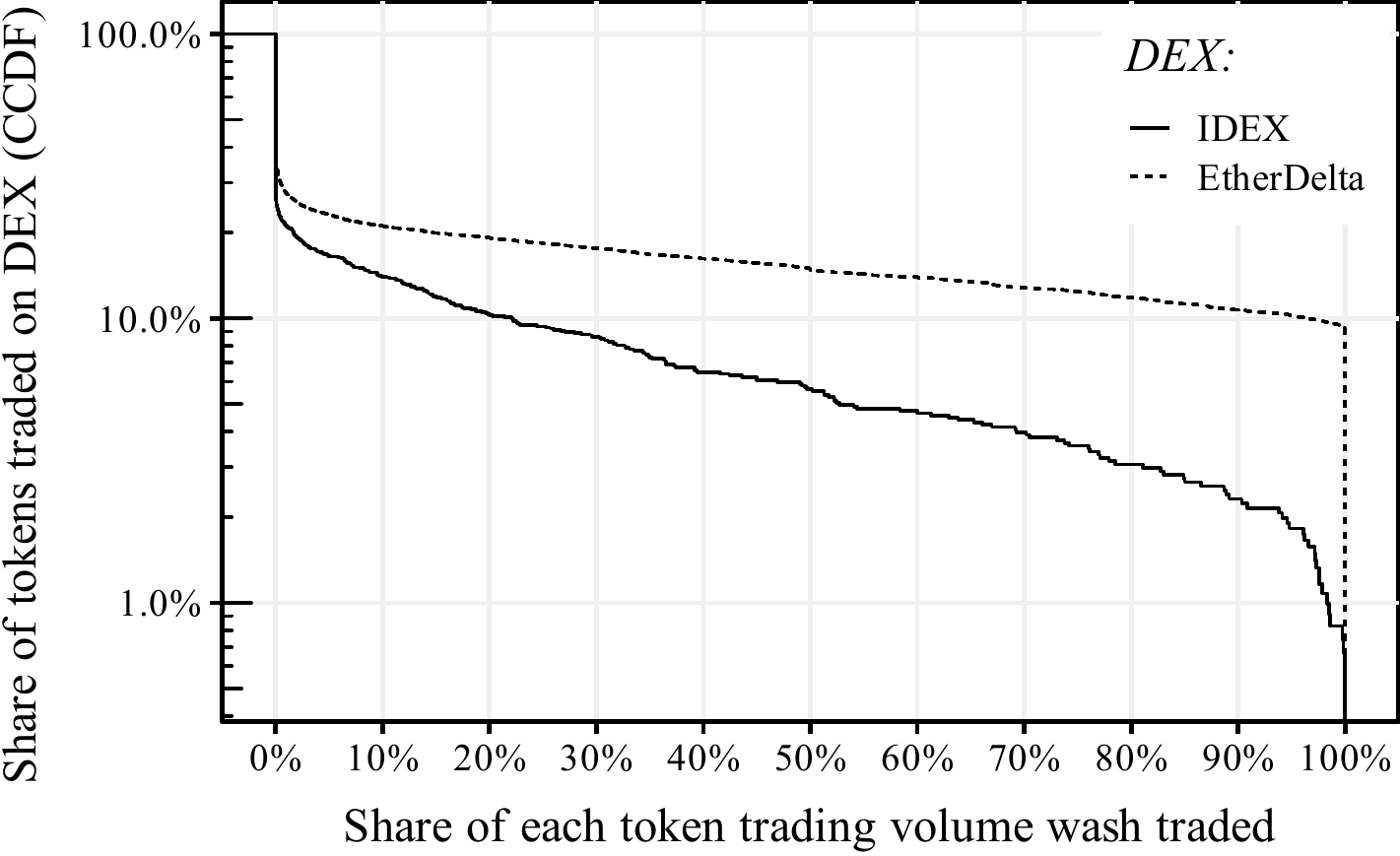}
    \caption{On both IDEX and EtherDelta, at least 10\% of the traded tokens have a total wash trading share of at least 20\%. On EtherDelta, there is the extreme case that 10\% of the tokens are entirely wash traded and have no real trading volume at all. On IDEX, this scenario only holds true for less than 1\% of the tokens.}
    \label{fig:token_wash_share}
\end{figure}

\subsection{Affected Tokens}
We now examine the extent to which the tokens traded on each DEX are affected by wash trading. We compute the share of a token's total trading volume that can be attributed to wash trading.
Based on this, we show the share of all tokens that exhibit at least a certain wash trading percentage in Figure~\ref{fig:token_wash_share}.
On both exchanges, at least 10\% of the traded tokens have a wash trading share of at least 20\%. On EtherDelta, almost 10\% of the tokens have been entirely wash traded. On IDEX, this only applies to less than 1\% of all traded tokens. While the figure illustrates the minimum share of wash trading activity, the total share of tokens that have ever been subject to at least one set of wash trades, irrespective of their volume, is 31.54\% for IDEX, and 42.12\% for EtherDelta.

\subsection{Wash Trade Activity in a Token's Lifespan}
\begin{figure}[ht]
    \centering
      \includegraphics[width=\linewidth]{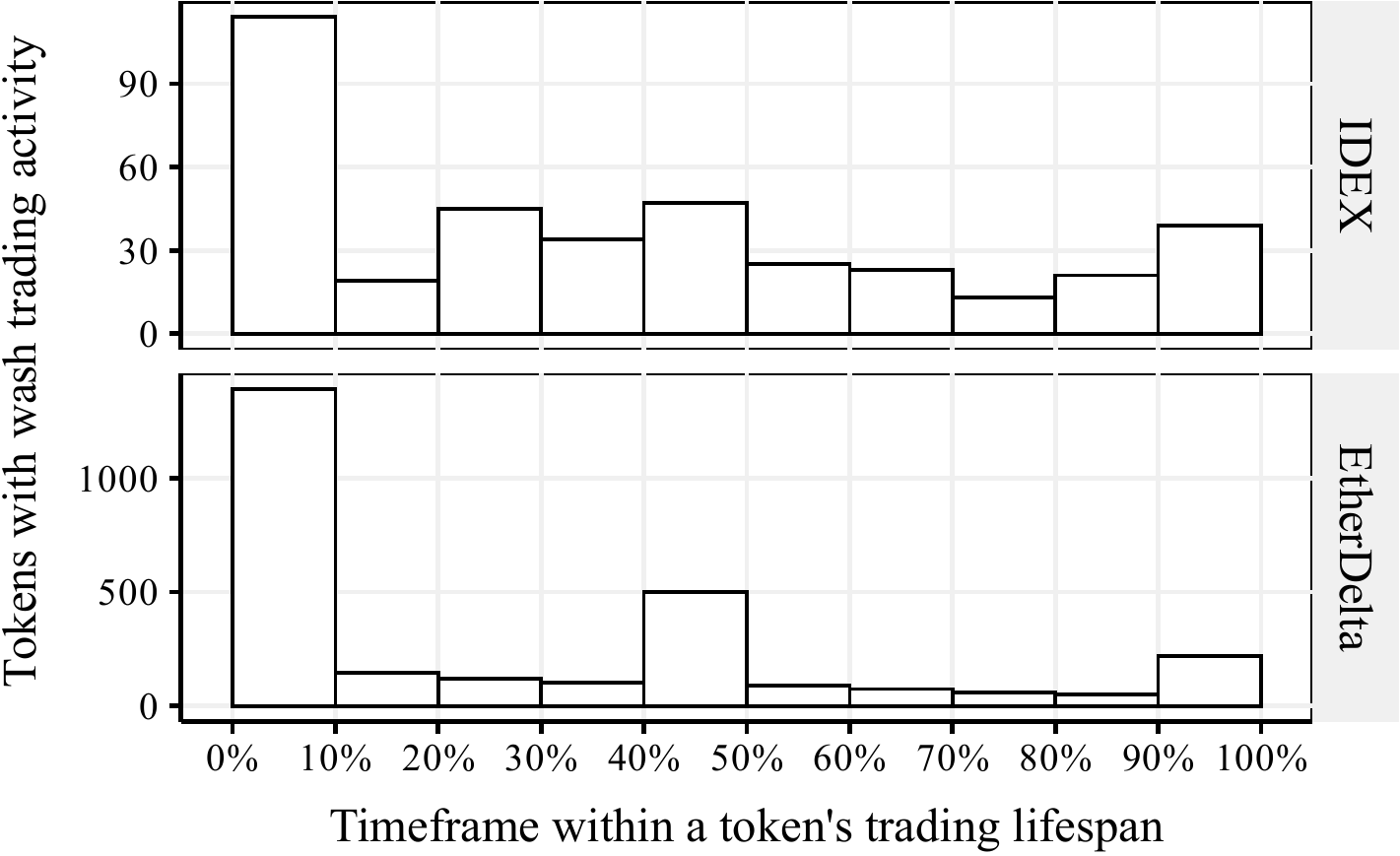}
    \caption{Distribution of median wash trading activity within a token's DEX trading lifespan. Frequently, a token is wash traded at the beginning of its trading lifespan, and at the very end. The peak at the center is due to some tokens that are entirely wash traded, where the median results in values close to 0.5.}
    \label{fig:token_wash_timeframes}
\end{figure}

To determine in which phase of a token's trading lifespan wash trading occurs, we first determine the relative time position of each wash trading activity in relation to the first and last (legitime) trading activity in the respective token. We then compute the median of these wash trading activites per token.
Figure~\ref{fig:token_wash_timeframes} shows the results for both exchanges. It can be seen that wash trading is more frequent at the beginning and at the end.
The peaks at the beginning of the tokens' lifespans may be attributed to manipulators wanting to generate trading volume so that real traders become aware of a new token.
The fact that wash trading also occurs frequently at the end could be due to a token showing only little trading activity and manipulators wanting to increase the trading activity again. Since the trading volume of 10\% of the tokens traded on EtherDelta consists almost exclusively of wash trades (see Figure~\ref{fig:token_wash_share}), the median wash trading time in the lifespan of these tokens may be close to 0.5. This may explain the peak of tokens with increased wash trading activities in the middle of the lifespan.

\subsection{Wash Trade Activity Over Time}

\begin{figure}[hb]
    \centering
      \includegraphics[width=\linewidth]{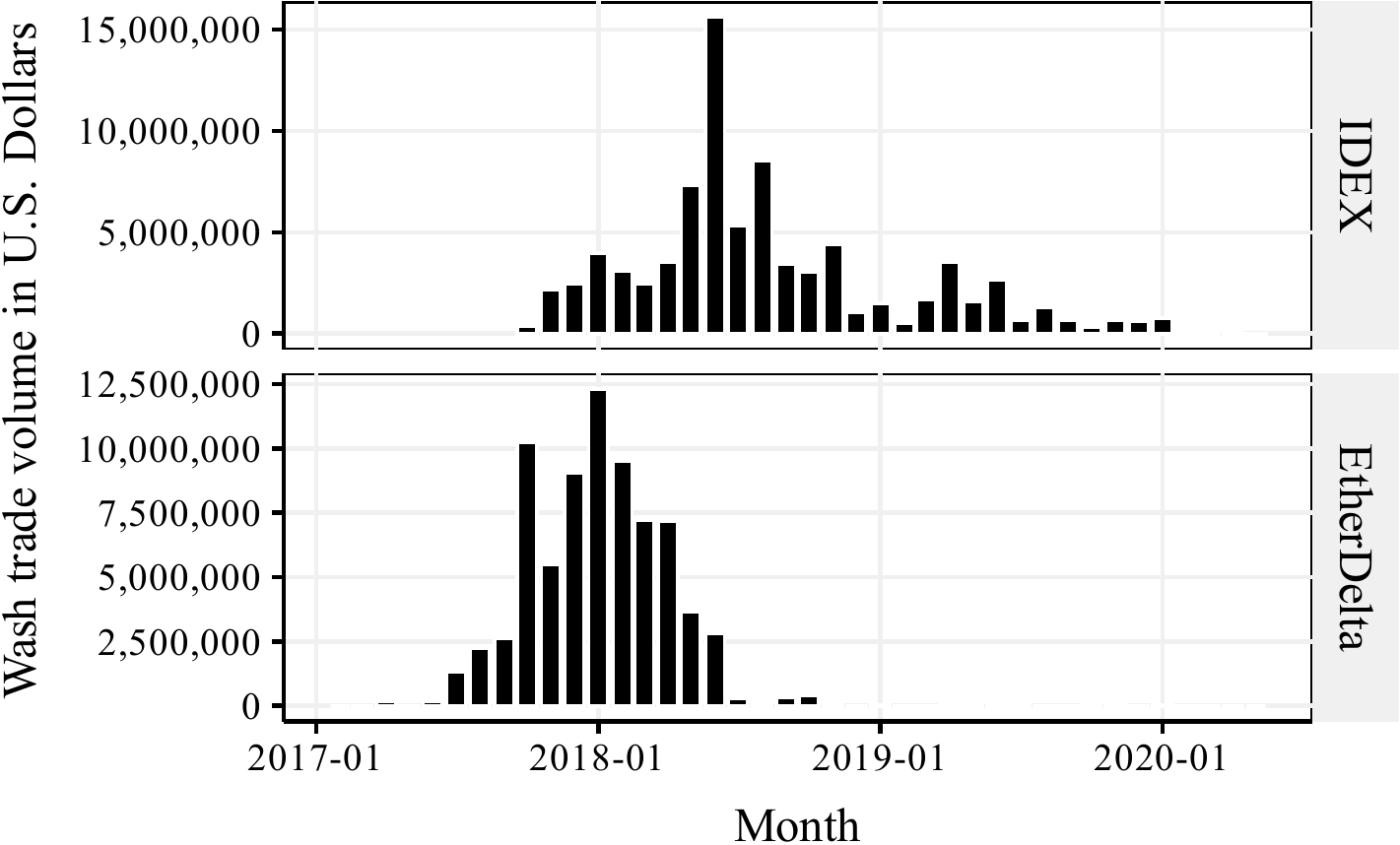}
    \caption{Absolute wash trading volume per month on both DEX. Most wash trading on EtherDelta occurred between mid-2017 and mid-2018. On IDEX, the activities started later and declined from the beginning of 2019.}
    \label{fig:monthly_wash_vol}
\end{figure}

We now look at the achieved wash trading volume over time, first from an absolute perspective and then relative to the rest of the (legitimate) trading volume. Figure~\ref{fig:monthly_wash_vol} shows the wash trading volume by month, over a period of about 3 years. IDEX activity started slightly later than on EtherDelta, reaching its peak in June 2018 with over \$15 million in monthly wash trading volume. On EtherDelta, activities are focused on the period from mid-2017 to mid-2018, with more than \$12 million in wash trading volume in January 2018. Activities on EtherDelta and IDEX flatten off sharply from mid-2018 and mid-2019 respectively, so that virtually no wash trading volume is visible.

\begin{figure}[ht]
    \centering
      \includegraphics[width=\linewidth]{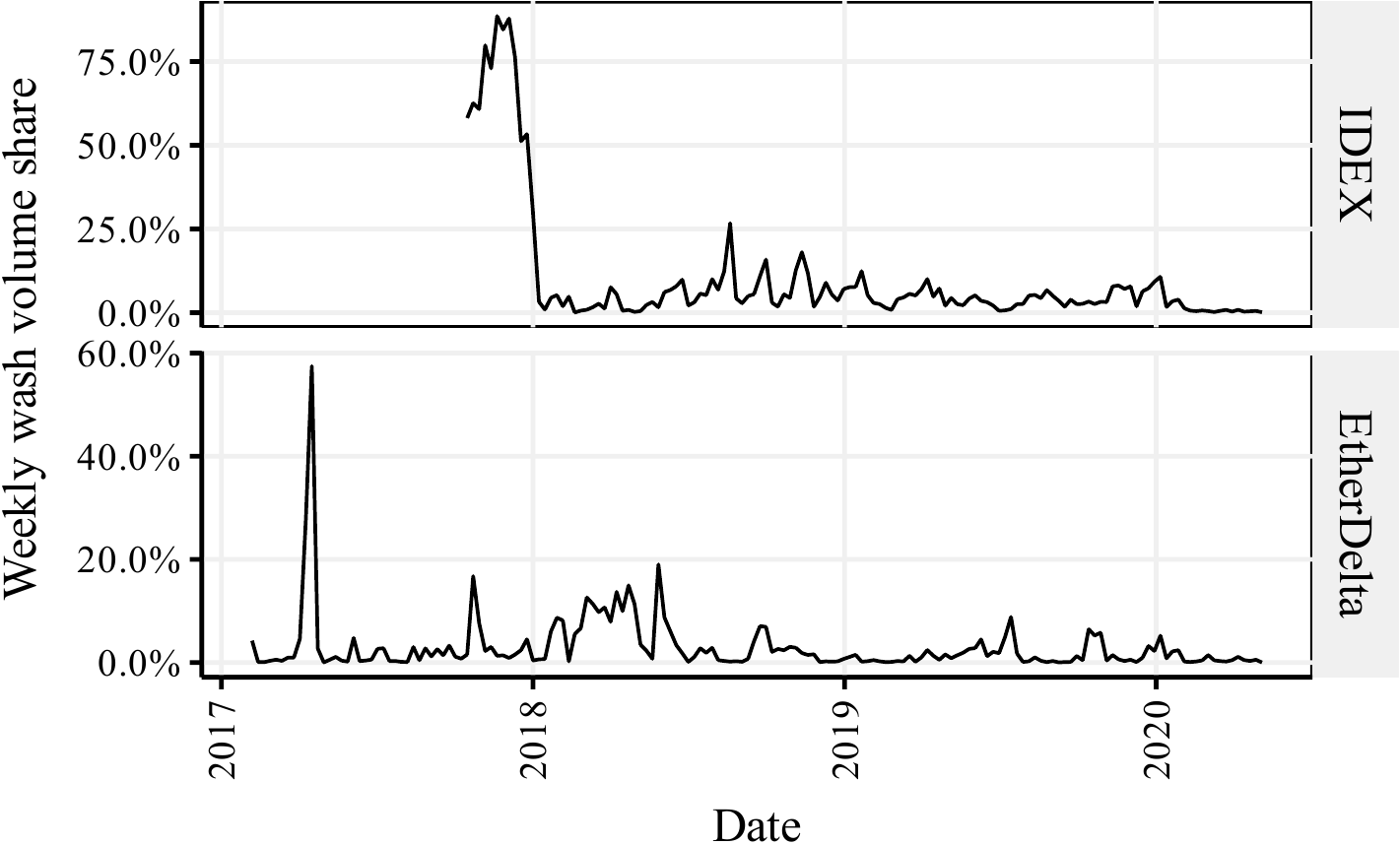}
    \caption{Weekly share of wash trade volume on both IDEX and EtherDelta. At the end of 2017, the majority of trades on IDEX were wash trades.}
    \label{fig:weekly_wash_vol_share}
\end{figure}

Figure~\ref{fig:weekly_wash_vol_share} shows the relative share of wash trading volume on a weekly basis.
Notably, IDEX exhibits several weeks with over 50\% and even close to 100\% of the total trading volume being wash traded at the end of 2017.
From 2018 onward, a share of up to 20\% can be observed on both exchanges, which then further declines in 2019 and 2020. %

Both graphs show peaks in late 2017 and early 2018. This falls into a phase in which the public awareness of cryptocurrencies and tokens was particularly high. At this time, many token prices reached record highs, only to be followed by a crash in early 2018.

\subsection{Summary}
Finally, we show detailed wash trading statistics for IDEX and EtherDelta in Table~\ref{tab:wash-trades-summary}.
Note that all statistics referring to wash trades include both self-trades and more complex wash trades.
On EtherDelta, the majority of wash trades are self-trades, whereas on IDEX, the self-trades only account for a small share. However, the total wash trading volume exceeds \$75 million on both exchanges, with a total of \ntotalWashTradingVolumeShort. On IDEX, illegitimate accounts traded an equivalent of over \$12 million with themselves, on EtherDelta almost \$65 million. The fraction of all trades that are wash trades is about 4\% on IDEX and 1.3\% on EtherDelta. On both exchanges, more than 30\% of the tokens that are tradable have been subject to wash trading activity.
On EtherDelta, the wash trades were executed by over 5,000 different accounts, leading to an average of about 13 wash trades per manipulative trader account. On the other hand, only 659 trader accounts were responsible for over 200,000 wash trades on IDEX.
Finally, considering that both DEX charge trading fees of 0.3\%, the total fees paid to the exchanges to perform wash trading, lead to an equivalent of 478,134 U.S. Dollars.

\begin{table}[!ht]
    \centering
    \caption{Wash Trades Summary for IDEX and EtherDelta.}
    \label{tab:wash-trades-summary}
    \begin{tabular}{| c | c | c |}
        \hline
        & \textbf{IDEX} & \textbf{EtherDelta} \\
        \hline
        \hline
        \# Self-Trades & \nIDEXSelfTrades & \nEtherDeltaSelfTrades \\
        \hline
        \# Wash Trades & \nIDEXWashTrades & \nEtherDeltaWashTrades \\
        \hline
        Self-Trades Share (Of All Trades) & \shareIDEXSelfTrades & \shareEtherDeltaSelfTrades \\
        \hline
        Wash Trades Share (Of All Trades) & \shareIDEXWashTrades & \shareEtherDeltaWashTrades\\
        \hline
        \hline
        Total Self-Traded Volume ETH & \totalIDEXSelfTradesETH & \totalEtherDeltaSelfTradesETH \\
        \hline
        Total Wash Volume ETH & \totalIDEXWashVolumeETH & \totalEtherDeltaWashVolumeETH\\
        \hline
        Total Self-Traded Volume USD & \totalIDEXSelfTradesUSD & \totalEtherDeltaSelfTradesUSD \\
        \hline
        Total Wash Volume USD & \totalIDEXWashVolumeUSD & \totalEtherDeltaWashVolumeUSD\\
        \hline
        Wash Trade Fees Received USD & 250,594 & 227,540 \\
        \hline
        \hline
        \# Self-Traded Tokens & \nIDEXSelfTokens & \nEtherDeltaSelfTokens \\
        \hline
        \# Wash Tokens & \nIDEXWashTokens & \nEtherDeltaWashTokens\\
        \hline
        Wash Token Share & \shareIDEXWashTokens & \shareEtherDeltaWashTokens\\
        \hline
        \hline
        \# Self-Trader Accounts & \nIDEXSelfTradeAccounts &\nEtherDeltaSelfTradeAccounts \\
        \hline
        \# Wash Trader Accounts & \nIDEXWashTradeAccounts & \nEtherDeltaWashTradeAccounts\\
        \hline
        \hline
        \# Analyzed SCCs & \nIDEXAnalyzedSCC & \nEtherDeltaAnalyzedSCC\\
        \hline
        \# SCCs with Wash Trading & \nIDEXWashSCC & \nEtherDeltaWashSCC\\
        \hline
        Mean \# Tokens Washed per SCC & \nIDEXMeanWashedTokenPerSCC & \nEtherDeltaMeanWashedTokenPerSCC \\
        \hline
    \end{tabular}
\end{table}

\section{Discussion}
Our wash trade detection method has identified various wash trading structures and manipulated trading volumes worth \$159 million. We have used several parameters for the method, which we have chosen very conservatively. This means our findings constitute a lower bound of the actual extent of wash trading activities. If we were to set these parameters in a more relaxed way, e.g. considering SCCs with at least $10$ repetitions instead of $100$, or a higher margin at which we define that a trader effectively did not change their position, the number of detected wash trades would be higher. At the same time it would be possible that some trades are falsely recognized as wash trades. An example of this would be individual traders who speculate in the short term, quickly buying back their sold tokens, while coincidentally trading with the same other trader. Especially in the case of self-trades, however, there is no room for interpretation.

The question of why so much wash trading is done at all can be answered as follows: Considering the thousands of different tokens that exist, it can be difficult for token founders to get enough attention to be listed on known exchanges. Therefore, some token founders may be thinking about giving their project a jump start with fake trading volume, which is one reason also suggested by a recent report~\cite{coindesk2019fakevolume}.
However, the wash trading activities on IDEX and EtherDelta have declined significantly since mid-2018. This is likely due to the emergence of new AMM-based exchanges such as Uniswap and others. We do believe that actors willing to boost a token using wash trades will also attempt to do so on the now more popular AMM DEX. The question remains how wash trading can be performed on these DEX, since users do not trade with each other but against a liquidity pool.

\subsection{Potential Countermeasures}
A standard countermeasure that is implemented by many exchanges in traditional markets is self-trade prevention functionality. A single account is prevented from filling its own buy or sell orders.
Even though IDEX's off-chain trade matching engine claims to have self-trade prevention functionality~\footnote{\url{https://docs.idex.io/\#self-trade-prevention}}, we have observed successful self-trades up until the end of our dataset (May 2020).
EtherDelta does not seem to implement such functionality.
However, as shown in the analysis of the wash trading structures, various other trading topologies exist that can circumvent such a self-trade prevention functionality.

While defining a limit to the number of trades that can be performed with the same trading partners might seem like a viable approach at first, it may disrupt legitimate trades, if the number of traders in a token market is relatively small. Another option to make it harder for wash traders to operate with multiple accounts is the introduction of Know Your Customer (KYC) procedures, that require traders to identify themselves with official identity information. This is standard in the traditional domain, and on most centralized exchanges, but not widely adopted in the realm of decentralized finance. IDEX has introduced KYC in July 2019\footnote{\url{https://medium.com/idex/idex-kyc-transition-period-and-updated-asset-availability-for-us-markets-set-to-begin-d45e945f842d}}, which coincides with lower wash trading volumes.

\section{Conclusion and Future Work}
This paper is the first to analyze the phenomenon of wash trading on decentralized cryptocurrency exchanges. We have presented a method to detect wash trades, and have examined the decentralized exchanges IDEX and EtherDelta on the Ethereum blockchain. On an empirical basis, we have identified wash trading activities in excess of 159 million U.S. Dollars and discovered common wash trading structures. These structures consist mainly of one or two accounts, but there also exist more complex patterns. Surprisingly, self-trades occur frequently, which could easily be prevented. On both exchanges, more than 30\% of all tokens have been subject to wash trading activity, and 10\% of the tokens on EtherDelta have been almost exclusively wash traded. These figures represent a lower bound, but yet underpin the need for countermeasures that also work in decentralized systems.

Future work can take multiple directions. Our volume matching algorithm could be extended to check all contiguous subsets, which would lead to approximately quadratic complexity, requiring a highly efficient implementation or distributed computing. In general, devising new wash trade detection algorithms can be of interest. These might also include temporality aspects, which we have not considered in this work.
We also consider the investigation of recently popularized DEX such as Uniswap to be promising, which are not based on the order book model. Furthermore, other aspects can be investigated, such as the effect of wash trades on price development and the token user community. Finally, new decentralized finance concepts such as loans, derivatives and insurances may offer further potential for manipulation which is not yet well understood.


\end{document}